\documentclass[useAMS,usenatbib]{mnras}
 
\usepackage{amsmath}
\usepackage{float}
\usepackage{graphicx}
\usepackage{txfonts}
\usepackage{indentfirst}
\usepackage{color}
\usepackage{ulem}
\usepackage{hyperref}
    
\newcommand{\eq}[1]{\begin{equation}  #1 \end{equation}}
\newcommand{\eqs}[1]{\begin{equation} \begin{split} #1 \end{split} \end{equation}}

\newcommand{\br}[1]{\left( #1 \right)}

\newcommand{\ba}[1]{\left\langle #1 \right\rangle}

\newcommand{\dd}{{\rm d}}
\newcommand{\expo}[1]{~{\rm e}^{ #1 }}
\newcommand{\vek}[1]{\mbox{\boldmath $#1$}}

\newcommand{\ic}{{i\mkern1mu}}
\newcommand{\svek}[1]{\mbox{\boldmath \scriptsize $#1$}}  

\def\araa{ARA\&A}
\def\apj{ApJ}
\def\apjl{ApJ}
\def\apjs{ApJS}
\def\aap{A\&A}

\def\mnras{MNRAS}
\def\nat{Nature}

\def\jgr{J.~Geophys.~Res}

\title[Turbulence in stratified ICM]{Multi-scale analysis of turbulence
evolution in the density stratified intracluster medium}

\author[Xun Shi, Daisuke Nagai and Erwin T. Lau]{Xun Shi$^{1}$\thanks{E-mail:
xun@mpa-garching.mpg.de}, 
Daisuke Nagai$^{2}$, Erwin T. Lau$^{2}$ \\
$^{1}$Max-Planck-Institut f\"ur Astrophysik,
Karl-Schwarzschild-Stra{\ss}e 1, D-85740 Garching bei M\"unchen, Germany\\
$^{2}$Department of Physics, Yale University, New Haven, CT 06520, U.S.A.
}
  
\begin{document}

\maketitle
  
\label{firstpage}

\begin{abstract}
The diffuse hot medium inside clusters of galaxies typically exhibits turbulent
motions whose amplitude increases with radius, as revealed by cosmological
hydrodynamical simulations. However, its physical origin remains unclear.
It could either be due to an excess injection of turbulence at large radii,
or faster turbulence dissipation at small radii. We investigate this by
studying the time evolution of turbulence in the intracluster medium (ICM) after
major mergers, using the Omega500 non-radiative hydrodynamical cosmological
simulations. By applying a novel wavelet analysis to study the
radial dependence of the ICM turbulence spectrum,
we discover that faster turbulence dissipation in the inner high density
regions leads to the increasing turbulence amplitude with radius. We also find
that the ICM turbulence at all radii decays in two phases after a major merger: an early fast
decay phase followed by a slow secular decay phase. 
The buoyancy effects resulting from the ICM density stratification becomes
increasingly important during turbulence decay, as revealed by a decreasing
turbulence Froude number $Fr \sim \mathcal{O}(1)$.
 Our results indicate that the stronger density stratification and
 smaller eddy turn-over time are the likely causes of the faster
 turbulence dissipation rate in the inner regions of the cluster. 
\end{abstract}


\section[]{Introduction}
Turbulent motions in the intracluster medium (ICM) are an essential piece of
puzzle in many astrophysical questions such as the non-thermal pressure and
cluster cosmology \citep{lau09, bat12, nelson14, nelson14b, shi14}, kinetic AGN
feedback \citep{banerjee14, yang16, zhangcy18}, solution to the cooling flow problem
\citep{dennis05,zhuravleva14,zhuravleva17}, galaxy - intracluster medium interaction \citep{elzant04b, kim07},
diffusion of chemical elements and heat
\citep{rus10}, generation of intergalactic magnetic fields \citep{car02,
ensslin06, sub06, iapichino08, cho14, beresnyak16, vazza18}, and acceleration of
cosmic rays and the cluster radio emissions \citep{fuj03, kang07, hallman11,
bru11, vazza11a}. 

In many of these contexts, an extended cluster volume with a large
radial range is involved. Thus, it is important to know how the turbulent gas
motions are distributed from the cluster center to the outskirts.
Observations of intracluster gas motions are so far limited to a
single measurement with the \textsl{Hitomi} satellite and a few 
indirect estimations based on pressure and density fluctuations inferred from
Sunyaev-Zel'dovich effect (SZ) and X-ray observations
\citep{churazov12,khatri16}. Our current knowledge is predominantly from
cosmological hydrodynamical simulations \citep[e.g.][]{lau09,
bat12, nelson14b}. These simulations consistently show an increasing fraction of
non-thermal pressure from the turbulent motions with radius owing to a
larger amplitude of gas motions at larger radii, at least outside the regions of
strong AGN influence.

What underlying physics leads to this radial distribution of ICM turbulence is
so far unknown.
As a first step, we would like to distinguish whether this radial dependence of
turbulence results from that of turbulence injection or that of
turbulence dissipation. \citet{shi14} postulated that the cause for this radial dependence is the faster
dissipation of ICM turbulence at small radii, whereas the injection of ICM
turbulence is rather independent of radius. Based on this assumption, they formulated an
analytical model of ICM non-thermal pressure which successfully captured its radial, redshift and accretion rate dependencies revealed by numerical simulations \citep{shi15,shi16}. This success motivates us to revisit their
assumption directly using hydrodynamical simulations: do we see radial
dependence of ICM turbulence dissipation? If yes, what is its physical
origin?

Even answering the first, seemingly simple question requires non-trivial
analysis tools.
We need on one hand multi-scale information to study turbulence evolution, and on
the other hand local information to study its radial dependence. Here, we perform a wavelet
analysis that is optimized for this purpose. Compared to traditional
multi-scale method i.e. a Fourier analysis which completely mixes motions at
different locations, the wavelet analysis cleanly separates scale and
location, and enables us to construct turbulence spectra at each spatial
location. Regarding the second question, we are interested in the physical
quantities in the ICM that are radial dependent and may affect turbulence
dissipation. An obvious candidate is the density stratification of the ICM. We
will introduce theoretical expectations on how density stratification influences
turbulence dissipation in the next section.

We shall limit our study to the main source of ICM turbulence stirring: merger
events during the process of large scale structure assembly
\citep[e.g.][]{ryu03,dolag05,iapichino11,nagai13,min14, miniati15, vazza17}. 
In particular, we study the ICM turbulence evolution during and after a major
merger which, when exists, plays a dominant role on the evolution of ICM motions
\citep{paul11, nelson12}.

\section{Role of density stratification in turbulence evolution}
\label{sec:stratify}
In the classical Kolmogorov picture of homogeneous turbulence, the eddy
turn-over time $t_{\rm eddy} = \ell / v_{\rm eddy}$ describes how fast
turbulence cascades from large to small scales.
As the turn-over time is shorter on smaller scales, its value on the large,
energy containing scales limits the rate of turbulence energy flow through the
length scales, and thus determines the time scale of turbulence dissipation. In
the absence of an external force, turbulence evolves merely under its own inertia.

A density stratification introduces an external force -- the buoyancy force.
For the ICM, the effect of this buoyancy force has been discussed
in the context of internal gravity waves \citep{kim07,rus11,zhur14,zhangcy18},
mass transport by buoyant bubbles \citep{pope10}, as well as buoyancy
instabilities in the presence of magnetic fields and / or cosmic ray heating \citep{balbus00,quataert08,sharma09}.

The same buoyancy effect alters the kinematics of turbulence by opening up
 a new channel of energy flow to and from the gravitational potential energy,
significantly complicating the Kolmogorov picture. Notably, the buoyancy force
introduces a new time scale characterised by the Brunt-V\"ais\"al\"a frequency 
\eq{
\label{eq:NBV}
N_{\rm BV} = \sqrt{-\frac{g}{\gamma} \frac{\dd \ln K}{\dd r}} \,,
}
the oscillation frequency of density perturbations in a stratified
medium, where $-g$ is the magnitude of gravitational acceleration, $K$ is gas
entropy, and $\gamma$ is the gas equation of state. 
When the buoyancy effect is strong, the Brunt-V\"ais\"al\"a frequency
dominates the turbulence evolution, as is known in ocean and atmosphere sciences
\citep[e.g.][]{ozmidov65, stillinger83, hopfinger87}.
In this regime, buoyancy prevents the turbulence eddies from overturning,
causing their final conversion to a quasi-2D turbulence and g-mode gravity
waves within a few Brunt-V\"ais\"al\"a periods.

The transition to the strongly buoyancy-influenced regime occurs when the Froude number of
the turbulence eddies 
\eq{
\label{eq:Fr}
Fr = \frac{v_{\rm eddy}}{N_{\rm BV} \ell} 
} 
is of order unity \citep{riley03}, i.e. when the Brunt-V\"ais\"al\"a frequency
is comparable to the eddy turn-over frequency $1/t_{\rm eddy}$, or
equivalently, when the buoyancy force is comparable to the inertial force of turbulence. 

Apparently, estimating the ICM turbulence Froude number is important for
evaluating the significance of the buoyancy effect on ICM turbulence evolution
and the radial distribution of ICM gas motions.
Note that while the Brunt-V\"ais\"al\"a frequency is determined solely by the
ICM equilibrium structure, namely its density, pressure and mass
distributions, the Froude number is associated with turbulence
eddies and depends on length scale, radius, as well as the stage of turbulence
evolution. Thus, a full quantification of the Froude number also requires a 
multi-scale, local analysis method such as the wavelet analysis.

\section{Simulation and analysis method}
\begin{figure}
\centering
    \includegraphics[width=.42\textwidth]{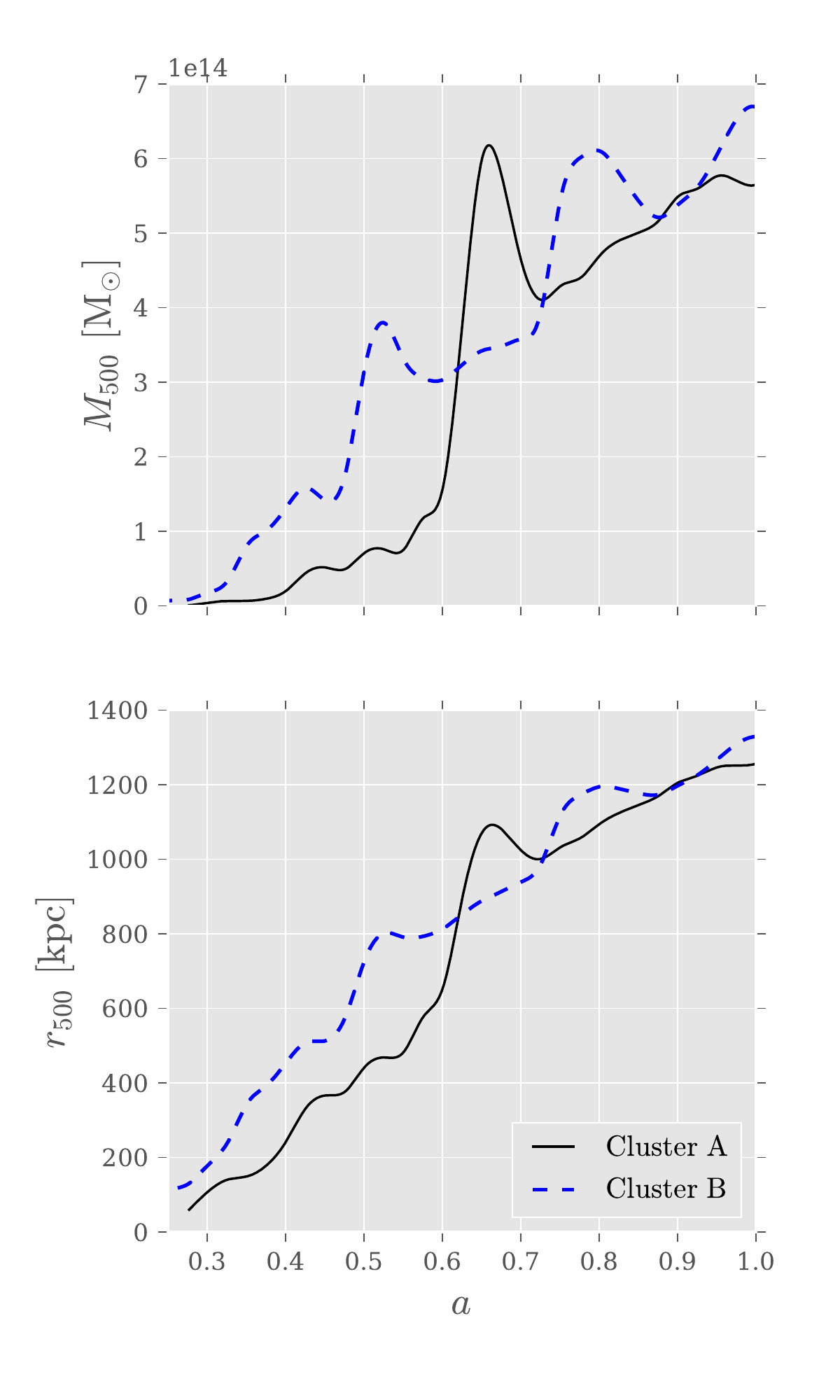}
  \caption{Mass accretion history (upper panel) and the growth of $r_{\rm 500}$
  of the selected clusters.}
\label{fig:mah}
\end{figure}

\begin{figure*}
\centering
    \includegraphics[width=0.85\textwidth]{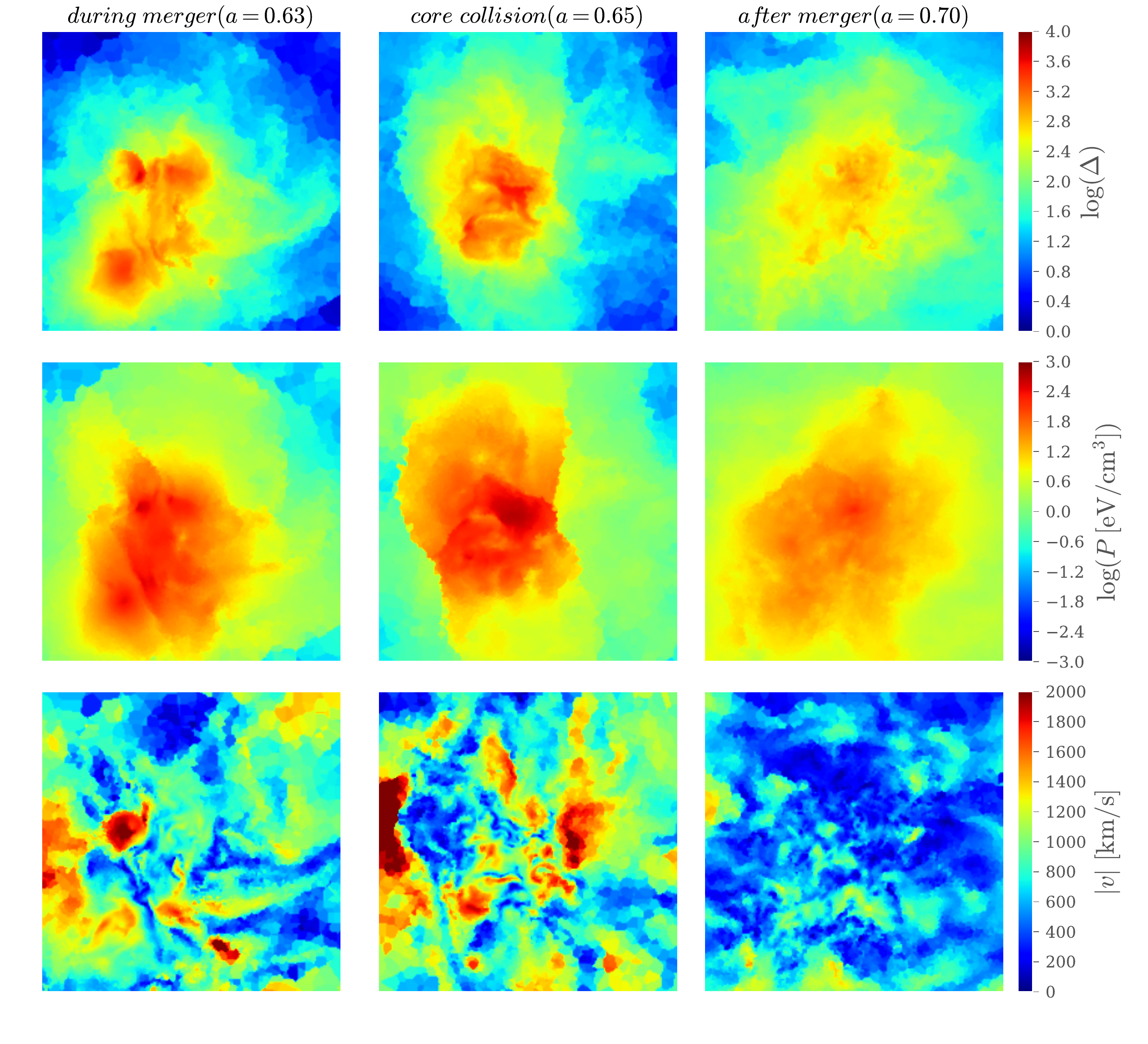}
    \caption{Overdensity, pressure and velocity amplitude (upper, middle and
    lower panels, respectively) of the central $r=1$ Mpc region of cluster A
    during its major merger epoch (left panels), at core collision of the merger (middle panels) and
    in the relaxation phase after the merger (right panels). Presented are
    $2$ Mpc by $2$ Mpc slices across the center with a thickness
    of $5$ kpc.}
\label{fig:image_ihalo2}
\end{figure*}

\subsection{Simulation and object selection}
We use the non-radiative runs of the Omega500 simulation \citep{nelson14},
a large cosmological Eulerian simulation performed with the Adaptive Refinement
Tree (ART) code \citep{krav99,krav02,rudd08} in a flat $\Lambda$CDM model with the \textsl{WMAP}
five-year cosmological parameters \citep{kom09}. The refinements give a
maximum comoving spatial resolution of $5.4$ kpc. The fine time resolution of the simulation
outputs ($0.02-0.03$ Gyr) allows us to follow the detailed time evolution of the gas motions. We refer to \citet{nelson14} for more detailed description
of the simulations.
 
We select one particular cluster (cluster A, see Figs.\;\ref{fig:mah} and
\ref{fig:image_ihalo2}) as our main object of study based on its special mass
assembly history. Cluster A gained most of its mass during an intense major
merger period between cosmic scale factors $a=0.55$ and $a=0.6$, with multiple
1:1 to 1:2 mergers within this short period of time. After the core collision which happens at $a=0.65$, it only had minor mergers with mass
ratios $<0.04$.
This unique mass assembly history allows a relatively clean study of the
injection of ICM gas motions by a major merger, and the subsequent dissipation
without much contamination from further injection.
As a comparison, we also examine cluster B, which has a similar mass at
redshift $z=0$ but has a much smoother mass accretion history
(Fig.\;\ref{fig:mah}). The most significant mergers experienced by cluster B were a 1:3 major merger at $a \approx
0.47$ and a 1:4 major merger at $a \approx 0.72$. It also has significant mass
contribution from more minor mergers.

We focus on the inner $r=1$ Mpc (physical) region of the clusters, which
corresponds roughly to $r_{\rm 500}$, a radius within which the average enclosed mass density is
$500$ times the critical density of the Universe, at redshift $0<z<1$ (see Fig.\;\ref{fig:mah}).
This limits the analysis to the virialized regions already at the major merger epoch.
To reduce the effect of non-uniform sampling from the cluster center to the
outskirts, we bin the simulated velocity field to a uniform resolution of $10$ kpc. 
This corresponds to a degradation of the resolution for most of the $r=1$ Mpc
region, except for its boundary where the resolution
sometimes drops to $\sim 20-30$ kpc. Therefore we limit our analysis to a
filtering scale $\ell\ge 28$ kpc and above. We also analyse a $5$ Mpc region with a lower resolution of
$50$ kpc to study the large scale modes.

\begin{figure}
\centering
    \includegraphics[width=0.48\textwidth]{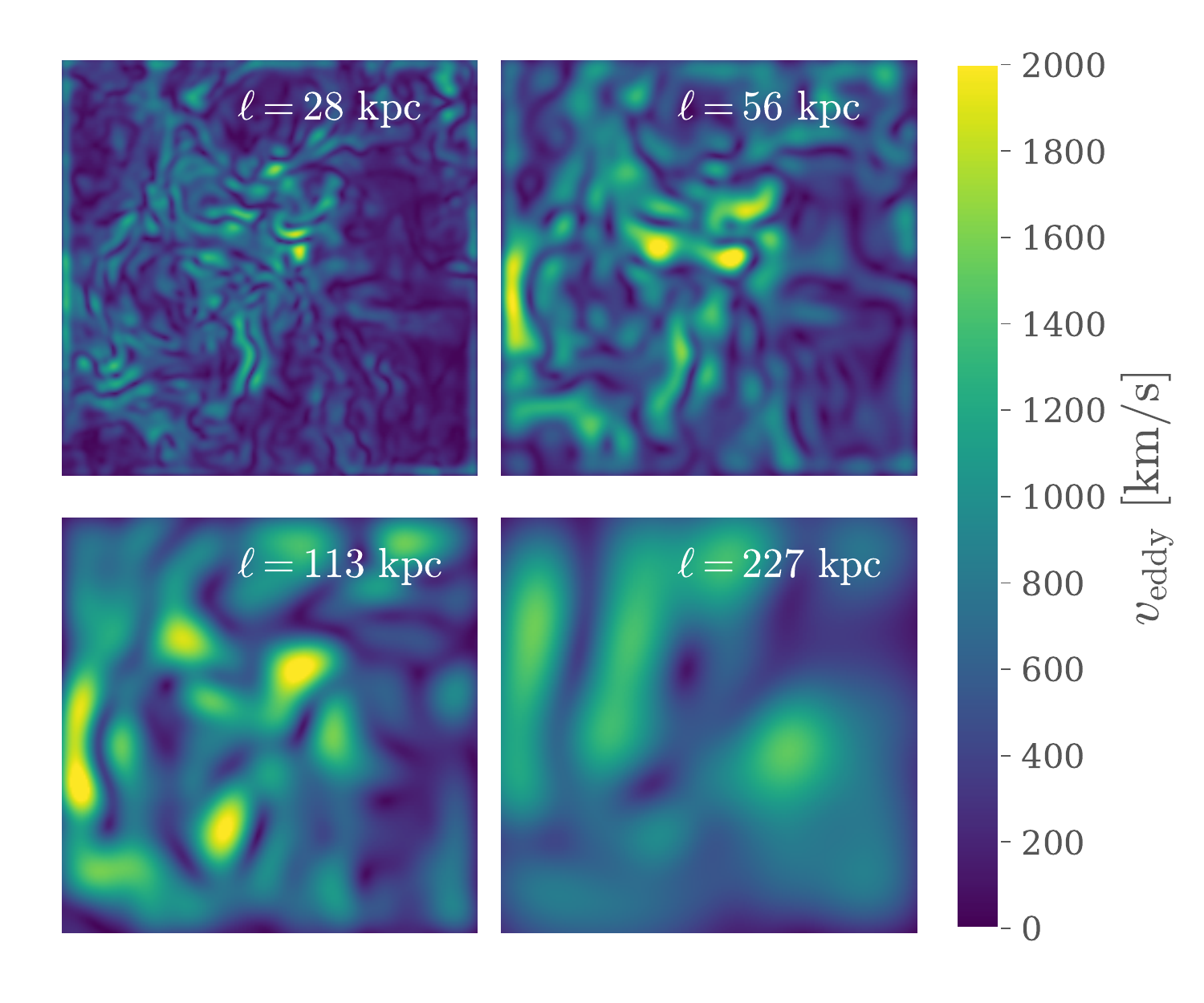}
    \caption{Wavelet decomposition of the velocity fields in the central $1$ Mpc
    region of cluster A at the core collision of its major merger epoch
    ($a=0.65$). Shown are slices of $v_{\rm eddy} = \sqrt{2 k
    E_{\rm wavelet}}$ at four different length scales. The slices are $2$ Mpc
    by $2$ Mpc  across the cluster center with a thickness of $5$
    kpc, same as in Fig.\;\ref{fig:image_ihalo2}.}
\label{fig:wavelet_ihalo2}
\end{figure}

\subsection{Wavelet analysis}
To study the role of stratification, it is necessary to distinguish two
lengths: turbulence eddy size and the cluster radius, and keep the cluster
radius as an individual variable in the analysis. One can in principle achieve this by
computing the Fourier power spectra in
radial bins. However, without homogeneity, the
infinite spatial support of the Fourier kernel $\expo{\ic \svek{k} \cdot
\svek{x}}$ becomes a disadvantage, and the physical significance of the Fourier
modes is significantly reduced. A different mode-decomposition basis with a
finite support in both Fourier-space and real-space, i.e. the wavelet, is
better suited for inhomogeneous data. Thus we perform a wavelet analysis to study the ICM velocity fields (Fig.\;\ref{fig:wavelet_ihalo2}, see Appendix.\;\ref{app:1} for detailed mathematics).
It corresponds to a filtering of the velocity field $\vek{v}(\vek{x})$
with a multi-scale filter function $\psi_{\ell, \svek{x}}$, resulting in scale-dependent signals
$\hat{\vek{F}}_{\ell}$ at each spatial location $\vek{x}$:
\eq{
\hat{\vek{F}}_{\ell}(\vek{x}) = \ba{\;\psi_{\ell, \svek{x}}\; ,\;
\vek{v}\;} \,, 
} 
where $\ba{}$ indicates the inner product. This enables us to 
construct a wavelet power spectrum 
\eq{
\label{eq:Ewavelet}
E_{\rm wavelet}(k)  =  \frac{ 32 \pi^4 k^2
|\hat{\vek{F}}_{\ell}(\vek{x})|^2}{2V}}
at each spatial location.
The wavelet power spectrum is a good substitute to the Fourier
velocity power spectrum $E(k)$. Compared to $E(k)$,  $E_{\rm wavelet}(k)$ can provide local spectrum
information, but it is associated with a filter function with a broader
$k-$space support and hence smoother than $E(k)$ in shape. For simplicity, we use a
symmetric Mexican-hat filter which ignores possible anisotropy, but the
compensated shape of the filter guarantees that large scale bulk motions do not create signals on small
scales.

Our wavelet approach is in contrast to the multi-scale filtering technique
used in \citet{vazza12}, which computes the average value of velocity around each
cell for increasingly larger scales by using top-hat filters of various sizes in
real space. 
The velocities filtered at different length scales can
also be used to construct local spectrum, although
the choice of the filters is less optimal for this purpose due to the
oscillatory shapes of their Fourier space correspondence (sinc functions).
However, rather than to obtain local spectrum information, \citet{vazza12}
designed the technique in order to find the largest correlation scale of the
local velocity field, which was then used to separate the velocity field
into `turbulence' and `bulk motions'. This classification of `turbulence' and
`bulk motions' is based on the idealised Kolmogorov picture of turbulence.
One potential issue with this approach lies in contamination by coherent
motions associated with small substructures and shocks.
To mitigate this contamination, we use the median instead of the mean for the
statistics, and use volume-weighted quantities to represent the volume-filling
medium. As shown by \citealt{zhu10, min14, vazza17} and \citealt{wittor17},
this volume-filling medium is dominated by solenoidal motions that are typical
for subsonic turbulence in the virial regions of galaxy clusters. 

\begin{figure*}
\centering
    \includegraphics[width=0.98\textwidth]{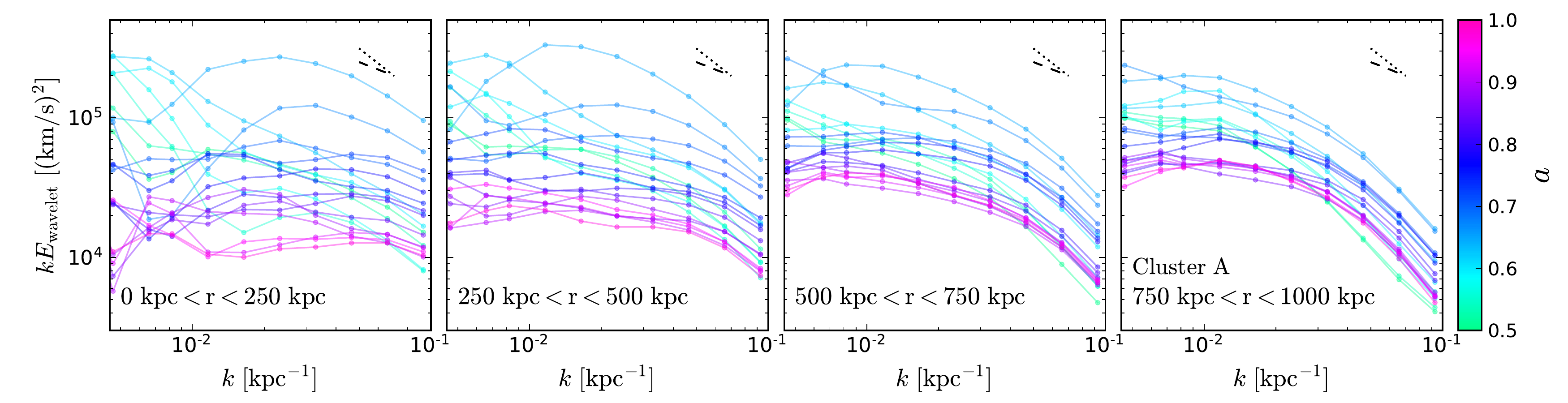}\\
    \includegraphics[width=0.98\textwidth]{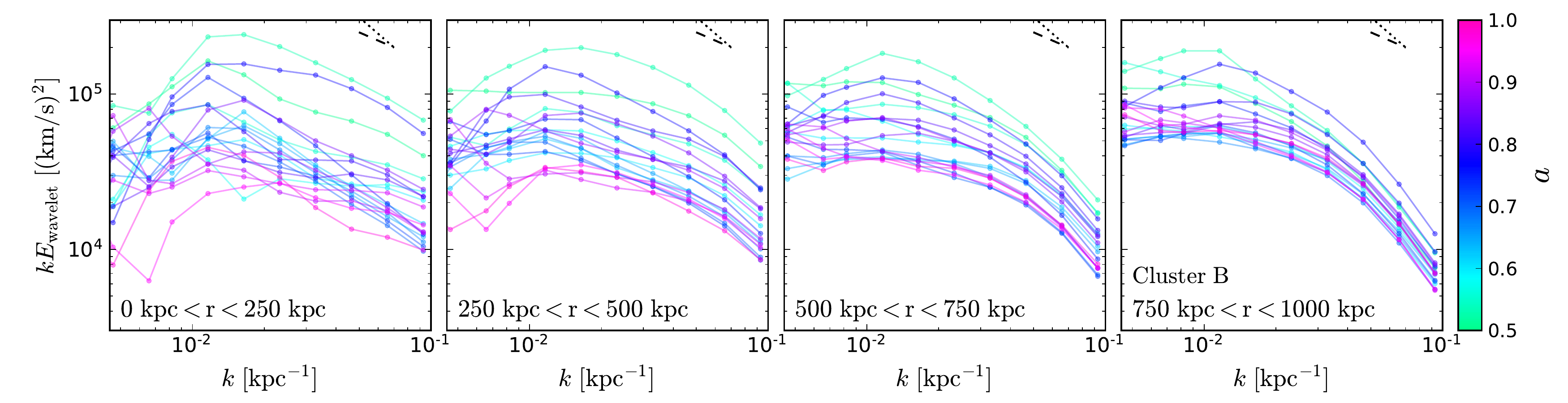}
    \caption{Wavelet power spectrum of ICM gas motions in clusters A (upper
    panel) and B (lower panel).
    Median wavelet power in four radial bins are shown from redshift one 
    till today (cosmic expansion factor $a$ is indicated by the color of the
    lines).
    For comparison, power-law slopes of $-2/3$ and $-4/3$ are shown on the upper
    right corner of each panel with dashed and dotted lines, with
    $-2/3$ being the value expected by the Komogorov scalings.
    The peak of the wavelet power spectrum $kE_{\rm wavelet}$ lies around $k
    \approx 0.01$ kpc$^{-1}$ between $0.5<a<1$, corresponding to a wavelet filter scale of $\ell = \sqrt{7/2}/k
\approx 200$ kpc (see Eq.\;\ref{eq:k2l}).
    }
\label{fig:kEk_A}
\end{figure*}

\section{Results}

\subsection{Radial-dependent evolution of ICM turbulence}
    
Given the ability to construct wavelet power spectrum $E_{\rm wavelet}$ 
(Eq.\;\ref{eq:Ewavelet}) at each spatial position, we can now study how the ICM
turbulence power spectra evolve with time as a function of radius. 
Fig.\;\ref{fig:kEk_A} shows the volume-weighted median
of the non-dimensional power spectra $k E_{\rm wavelet}$ in four radial bins for
cluster A and B. The results from the $1$ Mpc and the $5$ Mpc regions are combined to cover a larger
$k-$range. 

Fig.\;\ref{fig:kEk_A} shows the clear time evolution of the ICM turbulence. 
The maximum power occurs at the core collision of a major merger. For
cluster A, this happens at $a=0.65$, and for cluster B, this happens at
$a=0.51$, followed by another local maximum at $a=0.76$. Notably, at these
moments of maximum power, the non-dimensional power spectrum  $k E_{\rm wavelet}(k)$ peaks at
roughly the same value in different radial bins.
This indicates rather uniform injection of kinetic energy by the
major merger within the inner $1$ Mpc region, supporting \citet{shi14}'s
assumption that turbulence injection has little radial dependence. The
physical scales of the energy containing turbulence eddies are $\ell \approx
200$ kpc between $0.5<a<1$ (see Fig.\;\ref{fig:kEk_A}), which are much smaller
than $r_{\rm 500}$ of the clusters at these times.
 
After the core collision of a major merger, the ICM gas motions enter a
decaying epoch. From the upper panel of Fig.\;\ref{fig:kEk_A}, we can clearly
see that the decay is faster at smaller radii for cluster A. We expect the effect to be less
pronounced for cluster B which has more continuous injection of gas motions from its smoother mass accretion history.
Nevertheless, this faster decay in the central regions is
also evident for cluster B after both of its major merger events, as shown by
the lower panel of Fig.\;\ref{fig:kEk_A}. This is the central result of this
paper.

\subsection{Two-phase time evolution after major merger}
\label{sec:evo}

\begin{figure}
\centering
    \includegraphics[width=.4\textwidth]{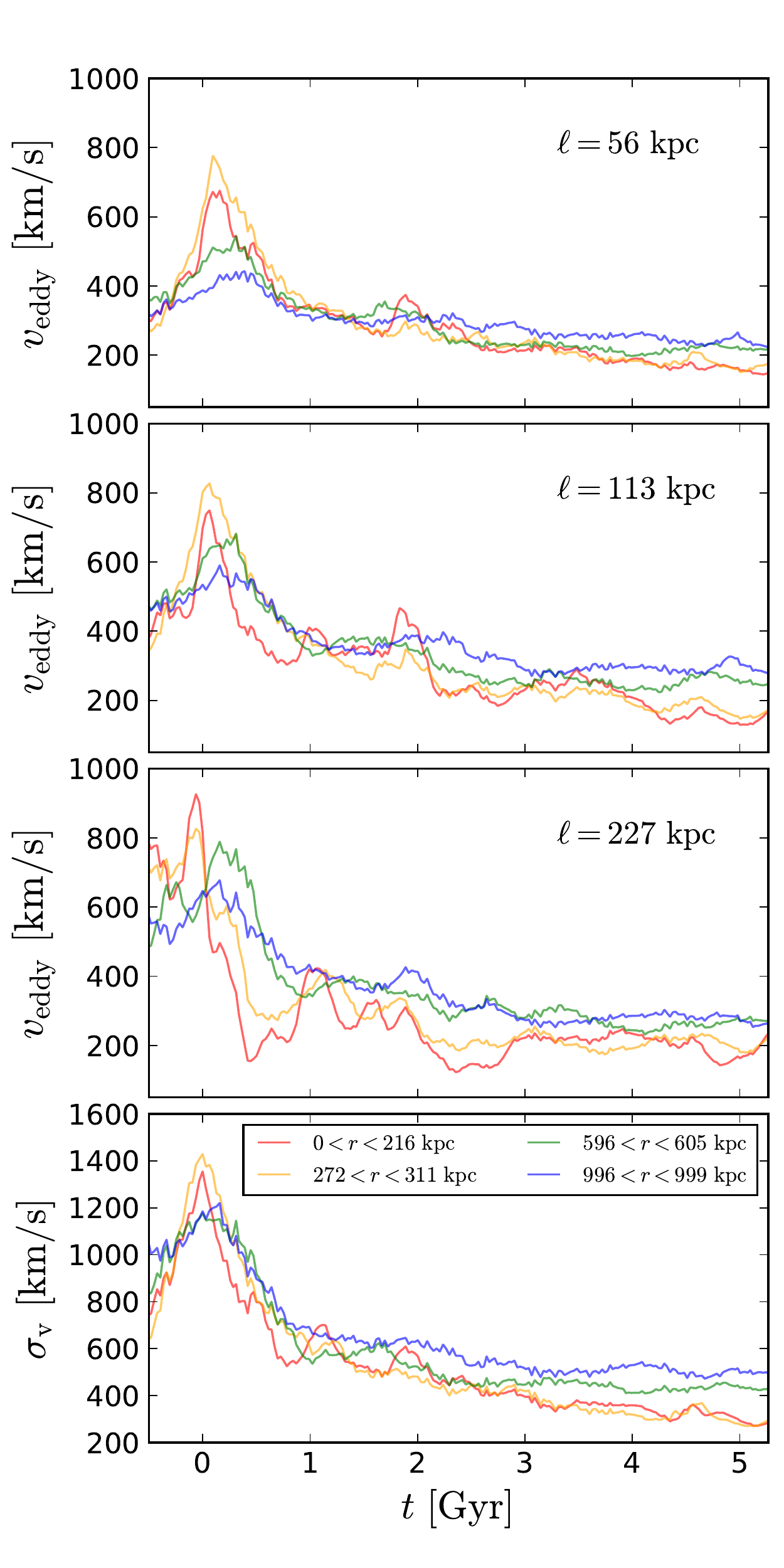} 
  \caption{Evolution of eddy velocities $v_{\rm eddy} = \sqrt{2 k E_{\rm 
  wavelet}}$ (top three panels for various physical scales) and $\sigma_{\rm
  v}$ (bottom panel) after the core collision at the major merger epoch
  ($t=0$) of cluster A. Here, volume-weighted median value is shown for
$v_{\rm eddy}$, and the value at $68.3$\% 
  of the total velocity pdf within the radial bin is used as $\sigma_{\rm v}$,
  in order to reduce sensitivity to the possible high velocity tails in the pdf
from shock fronts and substructures.
  Different colors represent different radial bins
  with bin sizes chosen such that they contain the same 3D volume.}
\label{fig:sigmav_wavelet_merger}
\end{figure}

\begin{figure}
\centering
    \includegraphics[width=.4\textwidth]{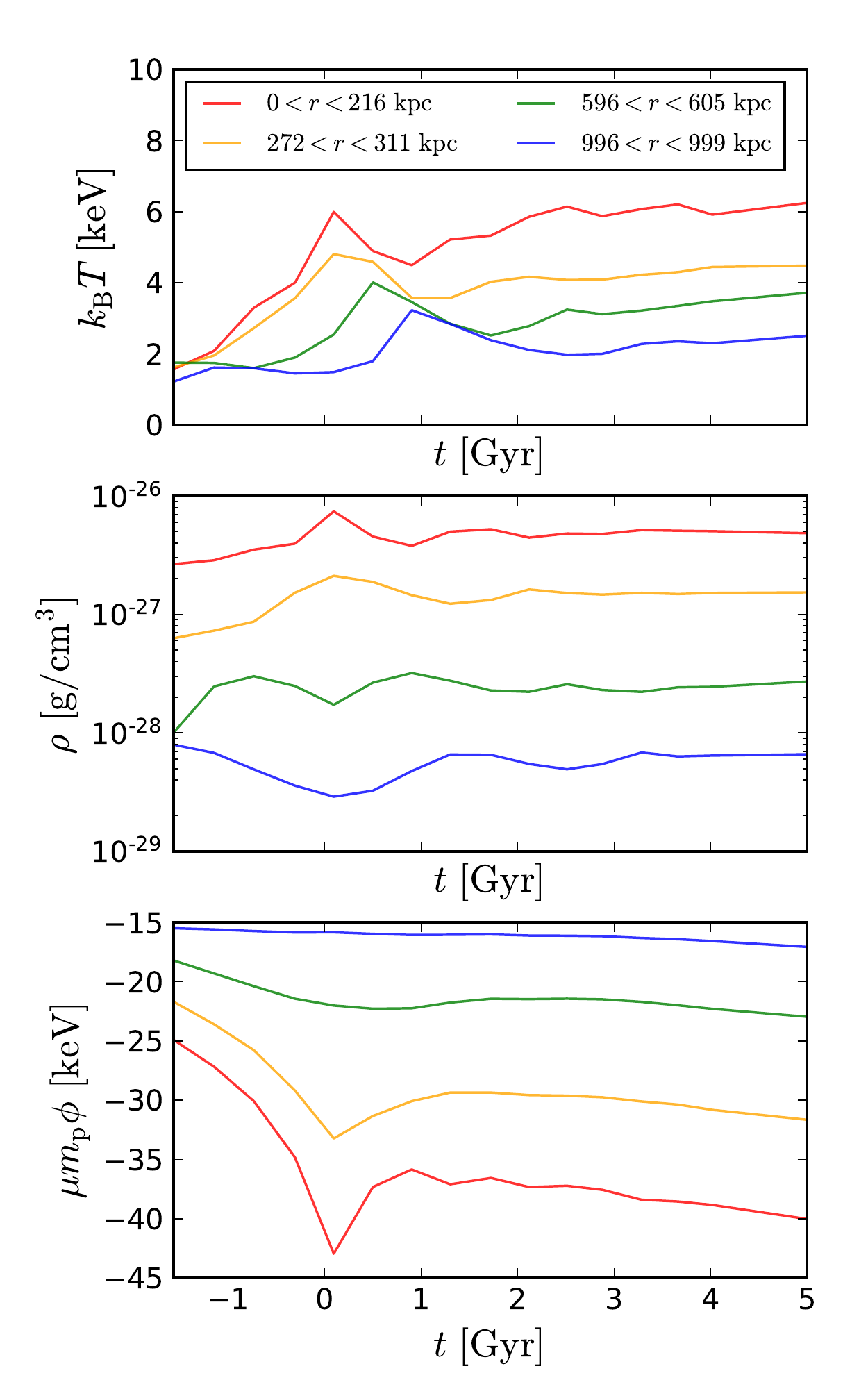}
    \caption{Evolution of gas temperature (top panel), density (middle panel)
    and gravitational potential (bottom panel) of cluster A since the beginning
    of the major merger epoch at $a=0.55$. Here, $k_{\rm B}$ is the Boltzmann constant, $\mu$ is the mean molecular weight,
    and $m_{\rm p}$ is the proton mass. In consistency with Figs.\;\ref{fig:sigmav_wavelet_merger}, $t=0$ marks the time of core collision and maximum kinetic energy density at $a=0.65$. }
\label{fig:heating}
\end{figure}

To further examine the decay of ICM gas motions after a major merger event, we
now focus on cluster A and examine the time evolution of velocities on
various length scales and at different radii
(Fig.\;\ref{fig:sigmav_wavelet_merger}). We see that the evolution of ICM
motions after a major merger event can be divided into two distinct phases: an
early fast decay phase of $\sim 1$ Gyr and a later secular decay phase. In
both phases, the decay of gas motions is faster in the central regions on all
physical scales between a few tens and a few hundreds kilo-parsecs (see also
Fig.\;\ref{fig:kEk_A}). 

The existence of the two decay phases points to complicated physics of ICM
turbulence dissipation after a major merger, involving more than one
dissipation time scales. To explore the
underlying physics of the two decay phases, we present the evolution of the 
gravitational potential, gas temperature and gas density in
Fig.\;\ref{fig:heating}. Remarkably, the time evolution of the ICM velocity
fields resembles that of the
gravitational potential rather than the gas temperature or density. In the time
evolution of the gas temperature and density, we can see clear propagation of the
peak temperature and density to larger radii after the core collision, which are
associated with the propagation of compression and rarefaction waves.
The evolution of the gravitational potential, on the other hand, is dominated by
that of the dark matter distribution, reflecting the collisionless passage of
dark matter halos. 
This resemblance between the time evolution of velocity fields and that of
gravitational potential indicates direct interaction between gravity and turbulence.
The peak positions of the gravitational potential and the velocity fields show
much less propagation, indicating a significant injection of turbulence
kinetic energy in the whole inner $1$ Mpc around the time of core collision. On
the other hand, the propagating compression and rarefaction waves have only a
minor effect on turbulence injection in comparison. Nevertheless, the crossing
time of these waves is similar to the duration of the dark matter halo
passage $\sim 1$ Gyr, which sets the duration of the first decay
phase.

One should bear in mind that the first decay phase, and the injection of
the turbulence, may depend on a merger configuration. Cluster A represents a
case where several proto-clusters merge from different directions within a
short period of time with small impact parameters, and the high density
cores are destroyed during the merger. We leave exploration of the dependence
on the merger configuration to future work.
 
The second decay phase of cluster A is not associated with significant density
variations, and is rather free of further merger events. In this phase, the
gravitational potential continues to steepen as a result of further dark matter
relaxation that lasts for a few Gyr \citep[e.g.][]{zhangcy16}. This
possibly injects more kinetic energy to the gas, albeit at a much smaller rate.
Interestingly, although the gravitational potential steepens more in the inner
regions, providing more source of kinetic energy there, the ICM velocities still
drop faster at smaller radii. We discuss plausible physical explanations
of this radial-dependent turbulence dissipation in the following section.

\begin{figure}
\centering
\includegraphics[width=.4\textwidth]{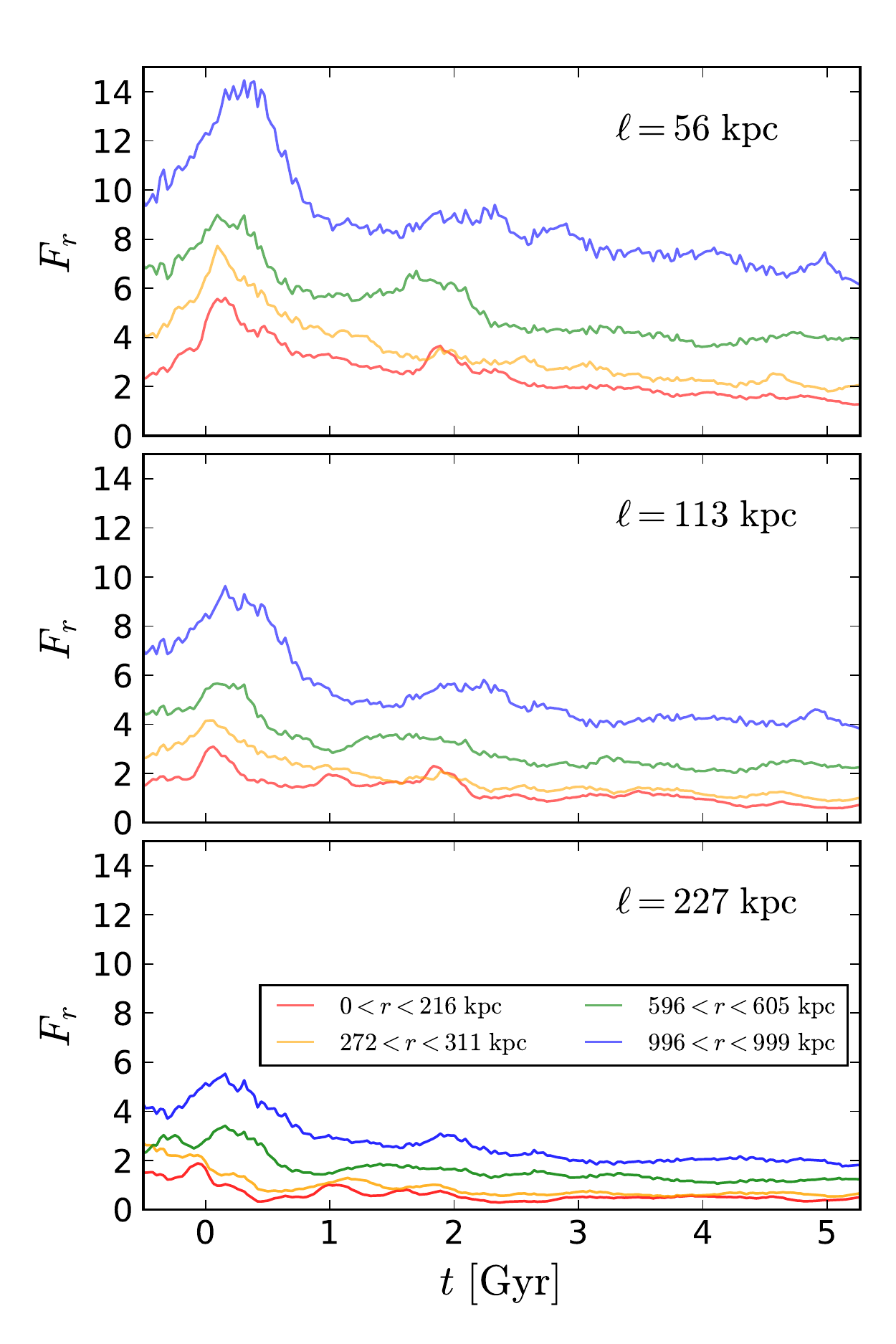}
  \caption{Evolution of the eddy Froude number as defined in Eq.\;\ref{eq:Fr}
  after the core collision at the major merger
  epoch ($t=0$) of cluster A. Different panels show different eddy sizes,
  and four radial bins are indicated with lines of different colors.}
\label{fig:Fr}
\end{figure}

\subsection{Physics of the radial-dependent turbulence dissipation}
For an isotropic, homogeneous turbulence described by the
Kolmogorov picture, the turbulence energy dissipates in an eddy turn-over time
$t_{\rm eddy}$. A radial-dependent $t_{\rm eddy}$ may explain the
radial-dependent turbulence dissipation found in the simulations. Namely,
when the turbulence energy of cluster A
and B peaks at major merger, $t_{\rm eddy}$ of the energy containing
turbulent eddies is longer in the outer regions mainly because of their
larger eddy sizes (Fig.\;\ref{fig:kEk_A}), causing a lower
dissipation rate there.
\citet{vazza17} estimated the local turbulence dissipation rate of a simulated
cluster immediately after its major merger using the Kolmogorov scaling
$\epsilon_{\rm diss} \propto v_{\rm eddy}^2 / t_{\rm eddy} \propto v_{\rm eddy}^3 / \ell$. They examined a much larger radial range
($r \approx 3 $ Mpc $\gtrsim 3 r_{\rm vir}$), and found a more
prominent decrease of the turbulence dissipation rate with radius (see their
Fig.\;13).
A caveat of such estimation, however, is that they should in principle be
applied only to turbulence eddies in the inertial range of the turbulence
spectrum. In practice, it is very difficult to identify and define the inertial 
range even in these
state-of-art simulations due to the still limited spatial resolution, as
reflected by the very limited range of the power spectrum slopes equaling the
$-5/3$ value expected for the inertial range (e.g. Fig.\;\ref{fig:kEk_A} of this paper, and
Fig.\;8 of \citealt{vazza17}).

As introduced in Sect.\;\ref{sec:stratify}, in a density-stratified medium like
the ICM, the Kolmogorov picture is incomplete and buoyancy strongly influences
turbulence dissipation when the turbulence Froude number is of order unity or
below. 
For a decaying turbulence, this condition on the Froude number is usually first
met by the largest eddies, as they have lower turn-over frequencies compared to
the smaller eddies. On the other hand, in a typical stratified
ICM, this condition is first met in the inner regions due to the stronger density
stratification characterised by a larger Brunt-V\"ais\"al\"a  frequency there.

Fig.\;\ref{fig:Fr} shows our wavelet-based evaluation of the eddy Froude number
$Fr$ for our cluster A. 
We find that $Fr$ is typically of order unity throughout the
evolution, except for small eddies ($\ell < 100$ kpc) in the outmost radial bin
during the first decay phase. This indicates that buoyancy is always
non-negligible on the energy containing turbulence eddies \footnote{However, since we define the Froude number $Fr$ using wavelet
filtered quantities $v_{\rm eddy}$ and $\ell$ as typical eddy velocity and size, matching these definitions to those used in the idealised
lab experiments may cause the critical value of $Fr$ to differ by a small
factor. This calibration requires dedicated simulation study which we leave to future work.}. As turbulence decays
further, smaller eddies, as well as eddies at larger radii are influenced more
strongly by buoyancy. The smallest Froude number occurs in the inner radial bins at late times, and reaches $Fr \sim 0.5$. 
This analysis supports the buoyancy time being an important underlying
time scale responsible for the turbulence decay in this phase. It also indicates
the importance of the energy channel to and from the gravitational potential
energy. Interestingly, this latter point also finds support from the
resemblance between the evolutions of $v_{\rm eddy}$ and $\phi$ in the fast decay phase
(Sect.\;\ref{sec:evo}). 

So far, our results are based on the non-radiative simulation. In the
presence of radiative cooling and feedback, the entropy gradient and
gravitational acceleration would be generally higher in the region of a
cool-core cluster. This would increase the Brunt-V\"ais\"al\"a frequency and
reduce the Froude number in the inner region, making the buoyancy effect more
pronounced there. Further investigation is required to reveal the detailed
physics of buoyancy influenced turbulence dissipation in the ICM.

\section{Conclusion and discussion}
We have investigated the physical origin of the spatial distribution of ICM
turbulence, whose amplitude increases with radius.
By applying a novel wavelet method, which enables a multi-scale, local analysis
of the ICM turbulence, we studied the injection of ICM turbulence at major
mergers and the subsequent turbulence dissipation in two galaxy clusters 
with distinctively different mass growth histories, extracted from the Omega500
non-radiative hydrodynamical cosmological simulation.
Our main results are summarized below:
 
\begin{enumerate}
\item
The peak kinetic energy injected into the ICM at a major merger event is nearly
independent of cluster radius, while turbulence at different radii decays at
different rate: faster in the central regions and slower in the outskirts. 

\item We found that the ICM turbulence decays in two distinct phases: an early
fast decay phase of $\sim 1$ Gyr associated with a rapid variation of the gravitational potential,
followed by slower, secular decay phase where the turbulence decays passively
in the stratified ICM.

\item A faster turbulence dissipation rate in the inner regions occurs
from a shorter eddy turn-over time, according to the Kolmogorov picture
of isotropic homogeneous turbulence. However, the turbulence
Froude number shows that the buoyancy effect resulting from ICM density
stratification is significant during the turbulence dissipation, indicating that
the Kolmogorov picture is inadequate for describing the ICM turbulence
evolution in the ICM.

\item  The Brunt-V\"ais\"al\"a buoyancy time scale is an important
time scale in ICM turbulence dissipation, and the energy channel to and from
gravitational potential energy is important in the evolution of ICM turbulence
energy.

\item Our results suggest that it is the combination of nearly uniform peak
kinetic energy at a major merger, and a slower decay of gas motions in the outer
regions, that leads to the increasing amplitude of ICM motions with
radius.  

\item Observationally, this can be tested by comparing mass biases of SZ and
X-ray cluster mass estimators constructed at different radii, and by constraining 
the ICM turbulence velocities using pressure and density fluctuations for clusters with spatially
resolved SZ/X-ray observations. 
Upcoming \textsl{XARM} and \textsl{Athena+} satellites equipped with a high-energy 
resolution calorimeter will likely provide direct constraints on the nature of the turbulence 
in the density stratified ICM.
\end{enumerate}

\section*{Acknowledgements}
This work is supported in part by NSF AST-1412768 and the facilities and staff
of the Yale Center for Research Computing. We thank the referee for his/her
helpful comments and suggestions.

\bibliographystyle{mnras}

\begin{thebibliography}{}
\makeatletter
\relax
\def\mn@urlcharsother{\let\do\@makeother \do\$\do\&\do\#\do\^\do\_\do\%\do\~}
\def\mn@doi{\begingroup\mn@urlcharsother \@ifnextchar [ {\mn@doi@}
  {\mn@doi@[]}}
\def\mn@doi@[#1]#2{\def\@tempa{#1}\ifx\@tempa\@empty \href
  {http://dx.doi.org/#2} {doi:#2}\else \href {http://dx.doi.org/#2} {#1}\fi
  \endgroup}
\def\mn@eprint#1#2{\mn@eprint@#1:#2::\@nil}
\def\mn@eprint@arXiv#1{\href {http://arxiv.org/abs/#1} {{\tt arXiv:#1}}}
\def\mn@eprint@dblp#1{\href {http://dblp.uni-trier.de/rec/bibtex/#1.xml}
  {dblp:#1}}
\def\mn@eprint@#1:#2:#3:#4\@nil{\def\@tempa {#1}\def\@tempb {#2}\def\@tempc
  {#3}\ifx \@tempc \@empty \let \@tempc \@tempb \let \@tempb \@tempa \fi \ifx
  \@tempb \@empty \def\@tempb {arXiv}\fi \@ifundefined
  {mn@eprint@\@tempb}{\@tempb:\@tempc}{\expandafter \expandafter \csname
  mn@eprint@\@tempb\endcsname \expandafter{\@tempc}}}

\bibitem[\protect\citeauthoryear{{Balbus}}{{Balbus}}{2000}]{balbus00}
{Balbus} S.~A.,  2000, \mn@doi [\apj] {10.1086/308732}, \href
  {http://adsabs.harvard.edu/abs/2000ApJ...534..420B} {534, 420}

\bibitem[\protect\citeauthoryear{{Banerjee} \& {Sharma}}{{Banerjee} \&
  {Sharma}}{2014}]{banerjee14}
{Banerjee} N.,  {Sharma} P.,  2014, \mn@doi [\mnras] {10.1093/mnras/stu1179},
  \href {http://adsabs.harvard.edu/abs/2014MNRAS.443..687B} {443, 687}

\bibitem[\protect\citeauthoryear{{Battaglia}, {Bond}, {Pfrommer}  \&
  {Sievers}}{{Battaglia} et~al.}{2012}]{bat12}
{Battaglia} N.,  {Bond} J.~R.,  {Pfrommer} C.,   {Sievers} J.~L.,  2012,
  \mn@doi [\apj] {10.1088/0004-637X/758/2/74}, \href
  {http://adsabs.harvard.edu/abs/2012ApJ...758...74B} {758, 74}

\bibitem[\protect\citeauthoryear{{Beresnyak} \& {Miniati}}{{Beresnyak} \&
  {Miniati}}{2016}]{beresnyak16}
{Beresnyak} A.,  {Miniati} F.,  2016, \mn@doi [\apj]
  {10.3847/0004-637X/817/2/127}, \href
  {http://adsabs.harvard.edu/abs/2016ApJ...817..127B} {817, 127}

\bibitem[\protect\citeauthoryear{{Brunetti} \& {Lazarian}}{{Brunetti} \&
  {Lazarian}}{2011}]{bru11}
{Brunetti} G.,  {Lazarian} A.,  2011, \mn@doi [\mnras]
  {10.1111/j.1365-2966.2010.17937.x}, \href
  {http://adsabs.harvard.edu/abs/2011MNRAS.412..817B} {412, 817}

\bibitem[\protect\citeauthoryear{{Carilli} \& {Taylor}}{{Carilli} \&
  {Taylor}}{2002}]{car02}
{Carilli} C.~L.,  {Taylor} G.~B.,  2002, \mn@doi [\araa]
  {10.1146/annurev.astro.40.060401.093852}, \href
  {http://adsabs.harvard.edu/abs/2002ARA%26A..40..319C} {40, 319}

\bibitem[\protect\citeauthoryear{{Cho}}{{Cho}}{2014}]{cho14}
{Cho} J.,  2014, \mn@doi [\apj] {10.1088/0004-637X/797/2/133}, \href
  {http://adsabs.harvard.edu/abs/2014ApJ...797..133C} {797, 133}

\bibitem[\protect\citeauthoryear{{Churazov} et~al.,}{{Churazov}
  et~al.}{2012}]{churazov12}
{Churazov} E.,  et~al., 2012, \mn@doi [\mnras]
  {10.1111/j.1365-2966.2011.20372.x}, \href
  {http://adsabs.harvard.edu/abs/2012MNRAS.421.1123C} {421, 1123}

\bibitem[\protect\citeauthoryear{{Dennis} \& {Chandran}}{{Dennis} \&
  {Chandran}}{2005}]{dennis05}
{Dennis} T.~J.,  {Chandran} B.~D.~G.,  2005, \mn@doi [\apj] {10.1086/427424},
  \href {http://adsabs.harvard.edu/abs/2005ApJ...622..205D} {622, 205}

\bibitem[\protect\citeauthoryear{{Dolag}, {Vazza}, {Brunetti}  \&
  {Tormen}}{{Dolag} et~al.}{2005}]{dolag05}
{Dolag} K.,  {Vazza} F.,  {Brunetti} G.,   {Tormen} G.,  2005, \mn@doi [\mnras]
  {10.1111/j.1365-2966.2005.09630.x}, \href
  {http://adsabs.harvard.edu/abs/2005MNRAS.364..753D} {364, 753}

\bibitem[\protect\citeauthoryear{{El-Zant}, {Kim}  \& {Kamionkowski}}{{El-Zant}
  et~al.}{2004}]{elzant04b}
{El-Zant} A.~A.,  {Kim} W.-T.,   {Kamionkowski} M.,  2004, \mn@doi [\mnras]
  {10.1111/j.1365-2966.2004.08175.x}, \href
  {http://adsabs.harvard.edu/abs/2004MNRAS.354..169E} {354, 169}

\bibitem[\protect\citeauthoryear{{En{\ss}lin} \& {Vogt}}{{En{\ss}lin} \&
  {Vogt}}{2006}]{ensslin06}
{En{\ss}lin} T.~A.,  {Vogt} C.,  2006, \mn@doi [\aap]
  {10.1051/0004-6361:20053518}, \href
  {http://adsabs.harvard.edu/abs/2006A%26A...453..447E} {453, 447}

\bibitem[\protect\citeauthoryear{{Farge}}{{Farge}}{1992}]{farge92}
{Farge} M.,  1992, \mn@doi [Annual Review of Fluid Mechanics]
  {10.1146/annurev.fl.24.010192.002143}, \href
  {http://adsabs.harvard.edu/abs/1992AnRFM..24..395F} {24, 395}

\bibitem[\protect\citeauthoryear{{Fujita}, {Takizawa}  \& {Sarazin}}{{Fujita}
  et~al.}{2003}]{fuj03}
{Fujita} Y.,  {Takizawa} M.,   {Sarazin} C.~L.,  2003, \mn@doi [\apj]
  {10.1086/345599}, \href {http://adsabs.harvard.edu/abs/2003ApJ...584..190F}
  {584, 190}

\bibitem[\protect\citeauthoryear{{Hallman} \& {Jeltema}}{{Hallman} \&
  {Jeltema}}{2011}]{hallman11}
{Hallman} E.~J.,  {Jeltema} T.~E.,  2011, \mn@doi [\mnras]
  {10.1111/j.1365-2966.2011.19637.x}, \href
  {http://adsabs.harvard.edu/abs/2011MNRAS.418.2467H} {418, 2467}

\bibitem[\protect\citeauthoryear{{Hopfinger}}{{Hopfinger}}{1987}]{hopfinger87}
{Hopfinger} E.~J.,  1987, \mn@doi [\jgr] {10.1029/JC092iC05p05287}, \href
  {http://adsabs.harvard.edu/abs/1987JGR....92.5287H} {92, 5287}

\bibitem[\protect\citeauthoryear{{Iapichino} \& {Niemeyer}}{{Iapichino} \&
  {Niemeyer}}{2008}]{iapichino08}
{Iapichino} L.,  {Niemeyer} J.~C.,  2008, \mn@doi [\mnras]
  {10.1111/j.1365-2966.2008.13518.x}, \href
  {http://adsabs.harvard.edu/abs/2008MNRAS.388.1089I} {388, 1089}

\bibitem[\protect\citeauthoryear{{Iapichino}, {Schmidt}, {Niemeyer}  \&
  {Merklein}}{{Iapichino} et~al.}{2011}]{iapichino11}
{Iapichino} L.,  {Schmidt} W.,  {Niemeyer} J.~C.,   {Merklein} J.,  2011,
  \mn@doi [\mnras] {10.1111/j.1365-2966.2011.18550.x}, \href
  {http://adsabs.harvard.edu/abs/2011MNRAS.414.2297I} {414, 2297}

\bibitem[\protect\citeauthoryear{{Kang}, {Ryu}, {Cen}  \& {Ostriker}}{{Kang}
  et~al.}{2007}]{kang07}
{Kang} H.,  {Ryu} D.,  {Cen} R.,   {Ostriker} J.~P.,  2007, \mn@doi [\apj]
  {10.1086/521717}, \href {http://adsabs.harvard.edu/abs/2007ApJ...669..729K}
  {669, 729}

\bibitem[\protect\citeauthoryear{{Khatri} \& {Gaspari}}{{Khatri} \&
  {Gaspari}}{2016}]{khatri16}
{Khatri} R.,  {Gaspari} M.,  2016, \mn@doi [\mnras] {10.1093/mnras/stw2027},
  \href {http://adsabs.harvard.edu/abs/2016MNRAS.463..655K} {463, 655}

\bibitem[\protect\citeauthoryear{{Kim}}{{Kim}}{2007}]{kim07}
{Kim} W.-T.,  2007, \mn@doi [\apjl] {10.1086/521950}, \href
  {http://adsabs.harvard.edu/abs/2007ApJ...667L...5K} {667, L5}

\bibitem[\protect\citeauthoryear{{Komatsu} et~al.,}{{Komatsu}
  et~al.}{2009}]{kom09}
{Komatsu} E.,  et~al., 2009, \mn@doi [\apjs] {10.1088/0067-0049/180/2/330},
  \href {http://adsabs.harvard.edu/abs/2009ApJS..180..330K} {180, 330}

\bibitem[\protect\citeauthoryear{{Kravtsov}}{{Kravtsov}}{1999}]{krav99}
{Kravtsov} A.~V.,  1999, PhD thesis, New Mexico State University

\bibitem[\protect\citeauthoryear{{Kravtsov}, {Klypin}  \& {Hoffman}}{{Kravtsov}
  et~al.}{2002}]{krav02}
{Kravtsov} A.~V.,  {Klypin} A.,   {Hoffman} Y.,  2002, \mn@doi [\apj]
  {10.1086/340046}, \href {http://adsabs.harvard.edu/abs/2002ApJ...571..563K}
  {571, 563}

\bibitem[\protect\citeauthoryear{{Lau}, {Kravtsov}  \& {Nagai}}{{Lau}
  et~al.}{2009}]{lau09}
{Lau} E.~T.,  {Kravtsov} A.~V.,   {Nagai} D.,  2009, \mn@doi [\apj]
  {10.1088/0004-637X/705/2/1129}, \href
  {http://adsabs.harvard.edu/abs/2009ApJ...705.1129L} {705, 1129}

\bibitem[\protect\citeauthoryear{{Miniati}}{{Miniati}}{2014}]{min14}
{Miniati} F.,  2014, \mn@doi [\apj] {10.1088/0004-637X/782/1/21}, \href
  {http://adsabs.harvard.edu/abs/2014ApJ...782...21M} {782, 21}

\bibitem[\protect\citeauthoryear{{Miniati} \& {Beresnyak}}{{Miniati} \&
  {Beresnyak}}{2015}]{miniati15}
{Miniati} F.,  {Beresnyak} A.,  2015, \mn@doi [\nat] {10.1038/nature14552},
  \href {http://adsabs.harvard.edu/abs/2015Natur.523...59M} {523, 59}

\bibitem[\protect\citeauthoryear{{Nagai}, {Lau}, {Avestruz}, {Nelson}  \&
  {Rudd}}{{Nagai} et~al.}{2013}]{nagai13}
{Nagai} D.,  {Lau} E.~T.,  {Avestruz} C.,  {Nelson} K.,   {Rudd} D.~H.,  2013,
  \mn@doi [\apj] {10.1088/0004-637X/777/2/137}, \href
  {http://adsabs.harvard.edu/abs/2013ApJ...777..137N} {777, 137}

\bibitem[\protect\citeauthoryear{{Nelson}, {Rudd}, {Shaw}  \& {Nagai}}{{Nelson}
  et~al.}{2012}]{nelson12}
{Nelson} K.,  {Rudd} D.~H.,  {Shaw} L.,   {Nagai} D.,  2012, \mn@doi [\apj]
  {10.1088/0004-637X/751/2/121}, \href
  {http://adsabs.harvard.edu/abs/2012ApJ...751..121N} {751, 121}

\bibitem[\protect\citeauthoryear{{Nelson}, {Lau}, {Nagai}, {Rudd}  \&
  {Yu}}{{Nelson} et~al.}{2014a}]{nelson14}
{Nelson} K.,  {Lau} E.~T.,  {Nagai} D.,  {Rudd} D.~H.,   {Yu} L.,  2014a,
  \mn@doi [\apj] {10.1088/0004-637X/782/2/107}, \href
  {http://adsabs.harvard.edu/abs/2014ApJ...782..107N} {782, 107}

\bibitem[\protect\citeauthoryear{{Nelson}, {Lau}  \& {Nagai}}{{Nelson}
  et~al.}{2014b}]{nelson14b}
{Nelson} K.,  {Lau} E.~T.,   {Nagai} D.,  2014b, \mn@doi [\apj]
  {10.1088/0004-637X/792/1/25}, \href
  {http://stacks.iop.org/0004-637X/792/i=1/a=25} {792, 25}

\bibitem[\protect\citeauthoryear{{Ozmidov}}{{Ozmidov}}{1965}]{ozmidov65}
{Ozmidov} R.~V.,  1965, Atmos. Oceanic Phys., 1, 861

\bibitem[\protect\citeauthoryear{{Paul}, {Iapichino}, {Miniati}, {Bagchi}  \&
  {Mannheim}}{{Paul} et~al.}{2011}]{paul11}
{Paul} S.,  {Iapichino} L.,  {Miniati} F.,  {Bagchi} J.,   {Mannheim} K.,
  2011, \mn@doi [\apj] {10.1088/0004-637X/726/1/17}, \href
  {http://adsabs.harvard.edu/abs/2011ApJ...726...17P} {726, 17}

\bibitem[\protect\citeauthoryear{{Pope}, {Babul}, {Pavlovski}, {Bower}  \&
  {Dotter}}{{Pope} et~al.}{2010}]{pope10}
{Pope} E.~C.~D.,  {Babul} A.,  {Pavlovski} G.,  {Bower} R.~G.,   {Dotter} A.,
  2010, \mn@doi [\mnras] {10.1111/j.1365-2966.2010.16816.x}, \href
  {http://adsabs.harvard.edu/abs/2010MNRAS.406.2023P} {406, 2023}

\bibitem[\protect\citeauthoryear{{Quataert}}{{Quataert}}{2008}]{quataert08}
{Quataert} E.,  2008, \mn@doi [\apj] {10.1086/525248}, \href
  {http://adsabs.harvard.edu/abs/2008ApJ...673..758Q} {673, 758}

\bibitem[\protect\citeauthoryear{{Riley} \& {deBruynKops}}{{Riley} \&
  {deBruynKops}}{2003}]{riley03}
{Riley} J.~J.,  {deBruynKops} S.~M.,  2003, \mn@doi [Physics of Fluids]
  {10.1063/1.1578077}, \href
  {http://adsabs.harvard.edu/abs/2003PhFl...15.2047R} {15, 2047}

\bibitem[\protect\citeauthoryear{{Rudd}, {Zentner}  \& {Kravtsov}}{{Rudd}
  et~al.}{2008}]{rudd08}
{Rudd} D.~H.,  {Zentner} A.~R.,   {Kravtsov} A.~V.,  2008, \mn@doi [\apj]
  {10.1086/523836}, \href {http://adsabs.harvard.edu/abs/2008ApJ...672...19R}
  {672, 19}

\bibitem[\protect\citeauthoryear{{Ruszkowski} \& {Oh}}{{Ruszkowski} \&
  {Oh}}{2010}]{rus10}
{Ruszkowski} M.,  {Oh} S.~P.,  2010, \mn@doi [\apj]
  {10.1088/0004-637X/713/2/1332}, \href
  {http://adsabs.harvard.edu/abs/2010ApJ...713.1332R} {713, 1332}

\bibitem[\protect\citeauthoryear{{Ruszkowski} \& {Oh}}{{Ruszkowski} \&
  {Oh}}{2011}]{rus11}
{Ruszkowski} M.,  {Oh} S.~P.,  2011, \mn@doi [\mnras]
  {10.1111/j.1365-2966.2011.18482.x}, \href
  {http://adsabs.harvard.edu/abs/2011MNRAS.414.1493R} {414, 1493}

\bibitem[\protect\citeauthoryear{{Ryu}, {Kang}, {Hallman}  \& {Jones}}{{Ryu}
  et~al.}{2003}]{ryu03}
{Ryu} D.,  {Kang} H.,  {Hallman} E.,   {Jones} T.~W.,  2003, \mn@doi [\apj]
  {10.1086/376723}, \href {http://adsabs.harvard.edu/abs/2003ApJ...593..599R}
  {593, 599}

\bibitem[\protect\citeauthoryear{{Sharma}, {Chandran}, {Quataert}  \&
  {Parrish}}{{Sharma} et~al.}{2009}]{sharma09}
{Sharma} P.,  {Chandran} B.~D.~G.,  {Quataert} E.,   {Parrish} I.~J.,  2009,
  \mn@doi [\apj] {10.1088/0004-637X/699/1/348}, \href
  {http://adsabs.harvard.edu/abs/2009ApJ...699..348S} {699, 348}

\bibitem[\protect\citeauthoryear{{Shi} \& {Komatsu}}{{Shi} \&
  {Komatsu}}{2014}]{shi14}
{Shi} X.,  {Komatsu} E.,  2014, \mn@doi [\mnras] {10.1093/mnras/stu858}, \href
  {http://adsabs.harvard.edu/abs/2014MNRAS.442..521S} {442, 521}

\bibitem[\protect\citeauthoryear{{Shi}, {Komatsu}, {Nelson}  \& {Nagai}}{{Shi}
  et~al.}{2015}]{shi15}
{Shi} X.,  {Komatsu} E.,  {Nelson} K.,   {Nagai} D.~S.,  2015, \mn@doi [\mnras]
  {10.1093/mnras/stv036}, \href
  {http://adsabs.harvard.edu/abs/2015MNRAS.448.1020S} {448, 1020}

\bibitem[\protect\citeauthoryear{{Shi}, {Komatsu}, {Nagai}  \& {Lau}}{{Shi}
  et~al.}{2016}]{shi16}
{Shi} X.,  {Komatsu} E.,  {Nagai} D.,   {Lau} E.~T.,  2016, \mn@doi [\mnras]
  {10.1093/mnras/stv2504}, \href
  {http://adsabs.harvard.edu/abs/2016MNRAS.455.2936S} {455, 2936}

\bibitem[\protect\citeauthoryear{Stillinger, Helland  \& Van~Atta}{Stillinger
  et~al.}{1983}]{stillinger83}
Stillinger D.~C.,  Helland K.~N.,   Van~Atta C.~W.,  1983, \mn@doi [Journal of
  Fluid Mechanics] {10.1017/S0022112083001251}, 131, 91–122

\bibitem[\protect\citeauthoryear{{Subramanian}, {Shukurov}  \&
  {Haugen}}{{Subramanian} et~al.}{2006}]{sub06}
{Subramanian} K.,  {Shukurov} A.,   {Haugen} N.~E.~L.,  2006, \mn@doi [\mnras]
  {10.1111/j.1365-2966.2006.09918.x}, \href
  {http://adsabs.harvard.edu/abs/2006MNRAS.366.1437S} {366, 1437}

\bibitem[\protect\citeauthoryear{{Torrence} \& {Compo}}{{Torrence} \&
  {Compo}}{1998}]{torrence98}
{Torrence} C.,  {Compo} G.~P.,  1998, \mn@doi [Bulletin of the American
  Meteorological Society] {10.1175/1520-0477(1998)079<0061:APGTWA>2.0.CO;2},
  \href {http://adsabs.harvard.edu/abs/1998BAMS...79...61T} {79, 61}

\bibitem[\protect\citeauthoryear{{Vazza}, {Brunetti}, {Gheller}, {Brunino}  \&
  {Br{\"u}ggen}}{{Vazza} et~al.}{2011}]{vazza11a}
{Vazza} F.,  {Brunetti} G.,  {Gheller} C.,  {Brunino} R.,   {Br{\"u}ggen} M.,
  2011, \mn@doi [\aap] {10.1051/0004-6361/201016015}, \href
  {http://adsabs.harvard.edu/abs/2011A%26A...529A..17V} {529, A17}

\bibitem[\protect\citeauthoryear{{Vazza}, {Roediger}  \& {Br{\"u}ggen}}{{Vazza}
  et~al.}{2012}]{vazza12}
{Vazza} F.,  {Roediger} E.,   {Br{\"u}ggen} M.,  2012, \mn@doi [\aap]
  {10.1051/0004-6361/201118688}, \href
  {http://adsabs.harvard.edu/abs/2012A%26A...544A.103V} {544, A103}

\bibitem[\protect\citeauthoryear{{Vazza}, {Jones}, {Br{\"u}ggen}, {Brunetti},
  {Gheller}, {Porter}  \& {Ryu}}{{Vazza} et~al.}{2017}]{vazza17}
{Vazza} F.,  {Jones} T.~W.,  {Br{\"u}ggen} M.,  {Brunetti} G.,  {Gheller} C.,
  {Porter} D.,   {Ryu} D.,  2017, \mn@doi [\mnras] {10.1093/mnras/stw2351},
  \href {http://adsabs.harvard.edu/abs/2017MNRAS.464..210V} {464, 210}

\bibitem[\protect\citeauthoryear{{Vazza}, {Brunetti}, {Br{\"u}ggen}  \&
  {Bonafede}}{{Vazza} et~al.}{2018}]{vazza18}
{Vazza} F.,  {Brunetti} G.,  {Br{\"u}ggen} M.,   {Bonafede} A.,  2018, \mn@doi
  [\mnras] {10.1093/mnras/stx2830}, \href
  {http://adsabs.harvard.edu/abs/2018MNRAS.474.1672V} {474, 1672}

\bibitem[\protect\citeauthoryear{{Wittor}, {Jones}, {Vazza}  \&
  {Br{\"u}ggen}}{{Wittor} et~al.}{2017}]{wittor17}
{Wittor} D.,  {Jones} T.,  {Vazza} F.,   {Br{\"u}ggen} M.,  2017, \mn@doi
  [\mnras] {10.1093/mnras/stx1769}, \href
  {http://adsabs.harvard.edu/abs/2017MNRAS.471.3212W} {471, 3212}

\bibitem[\protect\citeauthoryear{{Yang} \& {Reynolds}}{{Yang} \&
  {Reynolds}}{2016}]{yang16}
{Yang} H.-Y.~K.,  {Reynolds} C.~S.,  2016, \mn@doi [\apj]
  {10.3847/0004-637X/829/2/90}, \href
  {http://adsabs.harvard.edu/abs/2016ApJ...829...90Y} {829, 90}

\bibitem[\protect\citeauthoryear{{Zhang}, {Yu}  \& {Lu}}{{Zhang}
  et~al.}{2016}]{zhangcy16}
{Zhang} C.,  {Yu} Q.,   {Lu} Y.,  2016, \mn@doi [\apj]
  {10.3847/0004-637X/820/2/85}, \href
  {http://adsabs.harvard.edu/abs/2016ApJ...820...85Z} {820, 85}

\bibitem[\protect\citeauthoryear{{Zhang}, {Churazov}  \&
  {Schekochihin}}{{Zhang} et~al.}{2018}]{zhangcy18}
{Zhang} C.,  {Churazov} E.,   {Schekochihin} A.~A.,  2018, \mn@doi [\mnras]
  {10.1093/mnras/sty1269}, \href
  {http://adsabs.harvard.edu/abs/2018MNRAS.tmp.1212Z} {}

\bibitem[\protect\citeauthoryear{{Zhu}, {Feng}  \& {Fang}}{{Zhu}
  et~al.}{2010}]{zhu10}
{Zhu} W.,  {Feng} L.-l.,   {Fang} L.-Z.,  2010, \mn@doi [\apj]
  {10.1088/0004-637X/712/1/1}, \href
  {http://adsabs.harvard.edu/abs/2010ApJ...712....1Z} {712, 1}

\bibitem[\protect\citeauthoryear{{Zhuravleva} et~al.,}{{Zhuravleva}
  et~al.}{2014a}]{zhuravleva14}
{Zhuravleva} I.,  et~al., 2014a, \mn@doi [\nat] {10.1038/nature13830}, \href
  {http://adsabs.harvard.edu/abs/2014Natur.515...85Z} {515, 85}

\bibitem[\protect\citeauthoryear{{Zhuravleva} et~al.,}{{Zhuravleva}
  et~al.}{2014b}]{zhur14}
{Zhuravleva} I.,  et~al., 2014b, \mn@doi [\apjl] {10.1088/2041-8205/788/1/L13},
  \href {http://adsabs.harvard.edu/abs/2014ApJ...788L..13Z} {788, L13}

\bibitem[\protect\citeauthoryear{{Zhuravleva}, {Allen}, {Mantz}  \&
  {Werner}}{{Zhuravleva} et~al.}{2017}]{zhuravleva17}
{Zhuravleva} I.,  {Allen} S.~W.,  {Mantz} A.~B.,   {Werner} N.,  2017, arXiv:
  1707.02304, \href {http://adsabs.harvard.edu/abs/2017arXiv170702304Z} {}

\makeatother
\end{thebibliography}

\appendix

\section{Symmetric Mexican-hat wavelet transform in N-dimension}
\label{app:1}
Mexican hat wavelet, or the 2nd order Derivative of Gaussian wavelet, is a
real-valued isotropic wavelet basis function used for continuous wavelet transforms
 that can be easily extended in high-dimensions \citet{farge92}. 
Here we derive the forms of Mexican-hat wavelet in $N$ dimension. In the
analysis presented in this paper, $N=3$. 

The mother wavelet is, by definition,
\eq{
\label{eq:mexican_mother}
\phi(\vek{x}) = \alpha_N \br{N -
|\vek{x}|^2}
\exp{\br{-\frac{|\vek{x}|^2}{2}}}
}
with a prefactor 
\eq{
\alpha_N = \frac{2}{\sqrt{N(N+2)}} \pi^{-\frac{N}{4}} 
}
chosen such that the normalisation 
\eq{
 \int_{R^N} |\;\phi(\vek{x})\;|^2 \;\dd^N x  = 1 \,.
}
The Fourier transform of the mother wavelet is
\eqs{
\label{eq:mother_Fourier}
\tilde{\phi}(\vek{k}) & =  \int_{R^N}
\phi(\vek{x}) \expo{-\ic \svek{k} \cdot \svek{x}} \; \dd^N x \\
& = (2\pi)^{\frac{N}{2}} \alpha_N \; |\vek{k}|^2 
\exp{\br{-\frac{|\vek{k}|^2}{2}}} \,.
}

Then,
a family of translated, dilated wavelet functions can be constructed from the
mother wavelet as 
\eq{
\psi_{\ell, \svek{x'}}(\vek{x}) = \ell^{-\frac{N}{2}} 
\phi\br{\frac{\vek{x}-\vek{x'}}{\ell}} 
}
Here, the prefactor $\ell^{-N/2}$ is chosen to ensure 
\eq{
\int_{R^N} |\; \psi_{\ell, \svek{x'}}(\vek{x})\;|^2 \;\dd^N x
 =   \int_{R^N} |\; \phi(\vek{x})\;|^2 \;\dd^N
 x 
= 1 \,,
}
i.e. that the wavelet function at each scale $\ell$ is normalized to have unit
energy. The corresponding Fourier space wavelet functions are
\eqs{
\tilde{\psi}_{\ell, \svek{x'}}(\vek{k}) & =  \int_{R^N}
\psi_{\ell, \svek{x'}}(\vek{x}) \expo{-\ic \svek{k} \cdot \svek{x}} \; \dd^N x \\
& = \ell^{\frac{N}{2}} \;
  \tilde{\phi}(\ell \vek{k}) \expo{-\ic \svek{x'} \svek{k}} \,.
 }

Wavelet coefficients of a signal $f(\vek{x})$ are computed as the inner product
of the signal and the wavelet family
\eqs{
\label{eq:coefficients}
\hat{F}_{\ell}(\vek{x}) &= \ba{\;\psi_{\ell, \svek{x}}\; ,\; f\;} \\
& = \int_{R^N} f(\vek{x'}) \;\psi_{\ell, \svek{x}}(\vek{x'})\; \dd^N x' \\
& = \frac{1}{(2\pi)^N}\int_{R^N} \tilde{f}(\vek{k}) \; \tilde{\psi}_{\ell,
\svek{x}}(\vek{k})\; \dd^N k \,.}

The original signal $f(\vek{x})$ can be reconstructed from the wavelet
coefficients at location $\vek{x}$ only:
\eq{
f(\vek{x}) = \frac{1}{C_{\delta}} \int_0^{\infty} 
\hat{F}_{\ell}(\vek{x})  \frac{\dd \ell}{\ell^{1+N/2}}
}
where
\eq{
C_{\delta} :=  \frac{1}{\Omega_N} \int_{R^N} 
\tilde{\phi}(\vek{k}) \frac{\dd^N
\vek{k}}{|\vek{k}|^N} \,.
}
For Mexican-hat given by (\ref{eq:mexican_mother}), accounting for the maximum
$k_{\rm max}$ \citep[cf.][]{farge92}
\eq{
C_{\delta} = (2\pi)^{\frac{N}{2}} \alpha_N \br{1 - \expo{-\frac{k_{\rm
max}^2}{2}}} \,.
}

The 1d wavelet energy power spectrum of the signal $f$ is,
expressed using the wavelet coefficients (\ref{eq:coefficients}),
\eq{
\label{eq:Ewavelet2}
E_{\rm wavelet}(k)  =  \frac{(2\pi)^N \Omega_N k^{N-1}
|\hat{F}_{\ell}(\vek{x})|^2}{2V} \,, }
where
\eq{
\label{eq:k2l}
k =  \frac{\sqrt{2 + N/2}}{\ell}
}
is the equivalent Fourier frequency \citep[see e.g.][]{torrence98} of the
Mexican-hat wavelet (i.e. the Fourier frequency that maximizes the
non-dimensional wavelet power spectrum $k E_{\rm wavelet}(k)$ for $f(\vek{x}) =
\expo{\ic \svek{k}\cdot \svek{x}}$),
and $\Omega_N$ is the solid angle (i.e. the area of the unit sphere) in
N-dimension 
\eq{
\Omega_N = \frac{2 \pi^{\frac{N}{2}}}{\Gamma\br{\frac{N}{2}}} \,.
} 
Given the locality of the wavelet basis functions, the wavelet power
spectrum $E_{\rm wavelet}(k)$ can be constructed at each location, i.e. as a
function of $\vek{x}$. This enables studies of the spatial dependence of the signal, such
as the radial dependence of the ICM velocity field. In the limiting case of 
a homogeneous, isotropic velocity field $f=\vek{v}$, the wavelet power spectrum
averaged over the whole volume $\ba{E_{\rm wavelet}(k)}$ is a good substitute
to the velocity power spectrum obtained from Fourier transform 
\eq{
E(k) = \frac{(2\pi)^N \Omega_N  k^{N-1} |\tilde{v}(k)|^2 }{ 2V}  \,.
}
The difference between the two is that $\ba{E_{\rm wavelet}(k)}$ is
constructed with a broader k-space bandwidth as given by
Eq.\;\ref{eq:mother_Fourier}, rather than the infinitely narrow bandwidth given
by the Dirac delta function for $E(k)$. In this paper, instead, we take the average
over radial bins to respect the density-stratified nature of the ICM.

\end{document}